\documentclass[submitting]{nst}

\usepackage{subfigure,dcolumn}
\usepackage[T2A,T1]{fontenc}
\usepackage[russian,english]{babel}
\usepackage{comment}
\usepackage{float}
\usepackage{graphicx}
\usepackage{amssymb}
\usepackage{tikz-cd}
\usepackage{tikz}
\usepackage{xcolor}
\usetikzlibrary{arrows.meta}
\usetikzlibrary{arrows.meta,positioning}
\newcommand{\elbowarrow}{%
  \tikz[baseline=-0.55ex]{
    \draw[-{Stealth[length=1.5mm,width=1.2mm]}, line width=0.4pt]
      (0,1.3em) -- (0,0) -- (3.5em,0);
  }%
}
\usepackage{xspace}

\newcommand{\pT}{\ensuremath{p_\mathrm{T}}\xspace}

\usepackage{multirow}
\usepackage{listings}
\begin{document}

\title{Probing Quantum Numbers and Decay Branching Ratios of Exotic States via Entanglement‑Enabled Spin Interference}

\author{Di Zhang}
\email{dizhang@mail.sdu.edu.cn}\affiliation{Key Laboratory of Particle Physics and Particle Irradiation (MOE), Institute of Frontier and Interdisciplinary Science, Shandong University, Qingdao, China 266237}

\author{Zhangbu Xu}
\email{zxu22@kent.edu}
\affiliation{Physics Department, Kent State University, Kent, OH 44242, USA}
\affiliation{Physics Department, Brookhaven National Laboratory, Upton, NY 11973, USA}

\author{Chi Yang}
\email{chiyang@sdu.edu.cn}\affiliation{Key Laboratory of Particle Physics and Particle Irradiation (MOE), Institute of Frontier and Interdisciplinary Science, Shandong University, Qingdao, China 266237}

\begin{abstract}

Ultra-peripheral heavy-ion collisions (UPCs) coherently photo-produce vector mesons through two spatially separated and quantum-mechanically indistinguishable production sites, whose separation far exceeds the lifetime of the created resonance. The superposition of these amplitudes generates production-site entanglement, observed experimentally as entanglement-enabled spin-interference patterns in the angular distributions of the decay products. We show that these interference signatures provide a sensitive probe of spin-alignment transfer in hadronic decay chains, enabling intermediate-state quantum numbers and relative branching ratios to be measured from observed angular modulations. Using the decay $\rho(1450)\!\rightarrow\!\pi^{+}\pi^{-}\pi^{+}\pi^{-}$ as an example, we simulate the $a_{1}(1260)\pi$, $\rho(\pi\pi)_{S}$, and $\pi(1300)\pi$ channels and demonstrate that each produces a distinct azimuthal $\cos 2\phi$ modulation. The $\pi(1300)\pi$ mode shows a uniquely separated response, allowing its branching fraction to be extracted directly. These results establish production-site entanglement in UPCs as a selective tool for hadron spectroscopy, particularly for broad or overlapping resonances that are otherwise difficult to disentangle.

\end{abstract}


\maketitle

\section{Introduction}
Relativistic heavy ions generate extremely strong electromagnetic fields that are highly Lorentz contracted, compressing the fields into the transverse plane perpendicular to the beam direction. In the Equivalent Photon Approximation (EPA), originally introduced by Fermi~\cite{Fermi:1924tc} and later extended by von~Weizsäcker and Williams~\cite{vonWeizsacker:1934nji,Williams:1934ad}, these time-dependent electromagnetic fields can be represented as fluxes of linearly polarized quasi-real photons~\cite{Vidovic:1992ik,Brandenburg:2022tna}. 
Because the photon flux scales as $Z^{2}$, with Z denoting the nuclear charge, heavy‑ion beams provide an exceptionally intense source of quasi‑real photons, enabling high‑precision photonuclear and photon–photon studies ~\cite{Baur:2001jj,Brandenburg:2021lnj,Wang:2022ihj}.

Ultra-peripheral collisions (UPCs) occur when the impact parameter exceeds the sum of the nuclear radii, $b > R_{A} + R_{B}$, thereby suppressing hadronic interactions and ensuring the dominance of photon-induced processes~\cite{Kryshen:2013uba,Bertulani:2005ru,Baltz:2007kq}. Photon-induced reactions in UPCs comprise both photon–photon interactions and photonuclear interactions. Two-photon processes can produce lepton pairs, diphotons, and other final states, and play an important role in tests of quantum electrodynamics and in searches for new physics~\cite{Bertulani:2005ru,Brandenburg:2022tna,Wang:2022ihj,Brandenburg:2021lnj,Baltz:2007kq,Baur:2001jj,BAUR20071,Brandenburg:2025one}. In exclusive photonuclear interactions, a quasi-real photon fluctuates into a $q\bar q$ dipole that scatters diffractively from the target nucleus via Pomeron exchange, producing a vector meson such as $\rho^0$, $J/\psi$, or excited $\rho'$ states~\cite{Klein:2016yzr,STAR:2002caw,Brandenburg:2025one}. These characteristics make UPCs a uniquely versatile environment for studying strong‑interaction dynamics at high energies. Importantly, these exclusively produced vector mesons carry rich spin and polarization information, providing a natural bridge to the systematic study of spin phenomena in relativistic heavy-ion collisions.

In relativistic heavy-ion collisions, spin and polarization phenomena have been extensively observed and utilized as powerful tools to probe the fundamental properties of the interacting system, encompassing both hyperon global polarization and vector meson spin alignment~\cite{Liang:2004ph,Liang:2004xn,STAR:2017ckg,STAR:2022fan,Becattini:2024rev}. Extending beyond the strong interaction dynamics, the electromagnetic fields generated by the colliding nuclei also exhibit striking polarization effects. Over the past several years, extensive UPC measurements at the Relativistic Heavy Ion Collider (RHIC) and the Large Hadron Collider (LHC) have revealed a rich spectrum of photon‑induced phenomena. The STAR Collaboration reported the first observation of the Breit–Wheeler process, $\gamma\gamma \to e^{+}e^{-}$, in heavy-ion UPCs, reconstructing more than six thousand exclusive $e^{+}e^{-}$ pairs and observing a pronounced $\cos 4\phi$ modulation in the pair azimuthal angle~\cite{STAR:2019wlg}. The measured modulation matched QED expectations for collisions of linearly polarized photons~\cite{Li:2019yzy,Brandenburg:2022tna}, providing direct experimental confirmation of the strong linear polarization inherent to the photon fields in UPCs.

Additional experiments have explored photon‑induced processes in environments where hadronic interactions are present. ALICE identified excess $J/\psi$ production at very low transverse momentum in hadronic Pb--Pb collisions, attributing it to coherent photoproduction persisting from peripheral into semicentral events~\cite{ALICE:2022zso,ALICE:2015mzu}. ATLAS observed centrality-dependent broadening of acoplanarity and $k_{\perp}$ distributions in $\gamma\gamma \to \mu^{+}\mu^{-}$ production~\cite{ATLAS:2022yad}. At RHIC, STAR measured significant enhancements of low-\pT $e^{+}e^{-}$ pairs in non-central Au+Au and U+U collisions~\cite{STAR:2018ldd}, and later reported excess low-\pT $J/\psi$ yields in hadronic events~\cite{STAR:2019yox}. Collectively, these results show that coherent photon-induced processes remain experimentally visible even for impact parameters smaller than $2R_{A}$. Further insight was provided by CMS, which measured the acoplanarity of $\gamma\gamma \to \mu^{+}\mu^{-}$ pairs as a function of forward neutron multiplicity in Pb--Pb UPCs at $\sqrt{s_{\mathrm{NN}}}=5.02$~TeV, demonstrating that the transverse momentum of coherent photons increases as the impact parameter decreases~\cite{CMS:2020skx}.

A particularly striking feature of coherent photonuclear production is the interference between the two indistinguishable production amplitudes in heavy-ion UPCs~\cite{Zha:2018jin,Zha:2020cst,Xing:2020hwh}. Because a vector meson may be photoproduced coherently from either nucleus, the two amplitudes---originating from spatially separated production sites---remain quantum-mechanically indistinguishable. Their superposition introduces production-site entanglement, expressed through a relative phase $\exp(i\vec{P}\cdot\vec{b})$ that correlates the meson momentum $\vec{P}$ with the impact parameter $\vec{b}$~\cite{Zha:2020cst,Xing:2020hwh}. This entanglement produces measurable spin-interference patterns that encode the photon polarization, the impact-parameter direction, and the kinematics of the vector meson and its decay products. As a result, the spin information of the incident photons is transferred to experimentally accessible angular observables through the decay kinematics of the meson. The STAR Collaboration has directly observed these entanglement-enabled spin-interference effects in exclusive $\rho^{0}$ photoproduction using Au+Au, U+U, and p+Au collisions~\cite{STAR:2022wfe}, reporting a pronounced $\cos 2\phi$ modulation at low \pT in heavy-ion collisions but no significant modulation in p+Au. This behavior is consistent with strong interference arising from coherent production at the two entangled sites. Subsequent measurements further confirm this phenomenon in $J/\psi$ production at RHIC~\cite{STAR:2025wpi} and in $\rho$ photoproduction at the LHC~\cite{ALICE:2024ife}.

These advances demonstrate that UPCs function as a quantum-interferometric environment in which production-site entanglement and controlled photon polarization jointly generate experimentally accessible spin-interference signatures~\cite{Brandenburg:2025one,Brandenburg:2024ksp,Luo:2025ewj}. Related phenomena at the Electron-Ion Collider (EIC) have been actively investigated~\cite{Wu:2025dxg,Kesler:2025ksf,Blaizot:2025scr,Boer:2025ixc}. 

The observation of azimuthal modulations arising from spin interference in coherent vector-meson photoproduction motivates exploring whether such observables can provide complementary information for hadron spectroscopy. The quantum numbers of exotic hadrons are commonly extracted through amplitude or partial-wave analyses of their decay kinematics~\cite{Peters:2004qw,Salgado:2013dja,Ketzer:2019wmd}. Recent discovery of a Glueball-like candidate: $X(2370)$ with $J^{PC}=0^{-+}$, at BESIII~\cite{BESIIIGlueballPhysRevLett.132.181901} highlights the importance of determining the quantum numbers of the exotic hadron~\cite{tr1v-lly8,Huang:2025pyv}. 
In this work, we show that this entanglement-enabled spin-interference mechanism can be exploited as a selective probe of the quantum numbers and decay dynamics of resonant states produced in photonuclear interactions. The remaining sections develop this idea and, using $\rho(1450)\rightarrow\pi^{+}\pi^{-}\pi^{+}\pi^{-}$ as a case study, show how distinct spin‑interference patterns arising from different decay chains can be used to discriminate intermediate‑state quantum numbers and measure their relative branching fractions.

\section{Conceptual Framework}

In high-energy photonuclear interactions, a quasi-real photon can fluctuate into a quark--antiquark dipole that scatters elastically from the target nucleus, producing a vector meson with the same quantum numbers as the photon, $J^{PC}=1^{--}$. This mechanism naturally explains the photoproduction of the ground-state $\rho(770)$ as well as several excited $\rho$ states that have been experimentally observed~\cite{STAR:2002caw,STAR:2009giy,ALICE:2024kjy} and are summarized by the Particle Data Group. The most prominent excitations include $\rho(1450)$, $\rho(1700)$, and $\rho(2150)$. Spectroscopic studies suggest that $\rho(1450)$ is predominantly a $2\,{}^{3}S_{1}$ state~\cite{He:2013ttg,Close:1997dj}, while $\rho(1700)$ is a likely $1\,{}^{3}D_{1}$ state~\cite{He:2013ttg}. The nature of $\rho(2150)$ remains uncertain, with possible orbital angular momenta $L=0$ or $L=2$. In the four-pion final state, the $\rho(1450)$ contribution is significantly larger than that of higher-mass excitations~\cite{ALICE:2024kjy}, and the $S$-wave dominance of this state simplifies the theoretical treatment. For these reasons, the present work focuses on $\rho(1450)$ photoproduction in ultra-peripheral collisions.

Historically, the quantum numbers of mesonic resonances have been
determined through detailed partial-wave analyses of multi-hadron
final states. Such approaches have played a central role in
establishing the $J^{PC}$ assignments of light-meson excitations,
including the $\rho(1450)$ and related vector states, as demonstrated
by classical PWA studies such as
Gilman et al.~\cite{Gilman:1970vi}, Harari and Zarmi~\cite{Harari:1970fw},
and Bialas et al.~\cite{Bialas:1971xk}. More recently, high-statistics
amplitude analyses from experiments such as CRYSTAL BARREL have
provided refined constraints on intermediate-state quantum numbers
and decay couplings~\cite{Amsler1998,CRYSTALBARREL:2001ldq,Fix2024}. These methods form the
traditional foundation of hadron spectroscopy against which new
spin-interference–based quantum-number probes in ultra-peripheral
collisions can be compared. 
The decay channels of $\rho(1450)$ relevant to this analysis are listed in Table~\ref{tab:rho1450_decay_modes}, following the spectroscopy results of the CRYSTAL BARREL experiment~\cite{CRYSTALBARREL:2001ldq}. Because our observable is constructed from the $\pi^{+}\pi^{-}\pi^{+}\pi^{-}$ final state, the $\rho\rho$ and $h_{1}(1170)\pi$ decay modes are excluded from the signal model. In photonuclear production, the polarization of the incoming photon is transferred to the photoproduced vector meson to a good approximation under $s$-channel helicity conservation, particularly at small momentum transfer~\cite{Gilman:1970vi,Harari:1970fw,Bialas:1971xk}. This polarization is subsequently transmitted through the decay chain, enabling interference-sensitive angular observables.

The three principal decay chains considered in this work exhibit distinct mechanisms of spin transfer:
\begin{itemize}
    \item \textbf{Mode 1:} $\rho(1450)\to a_{1}(1260)\pi$. The parent spin aligns the $a_{1}(1260)$, which transfers it to the daughter $\rho$ via $a_{1}\to\rho\pi$, and finally to the orbital angular momentum (OAM) of the pions in $\rho\to\pi\pi$.
    \item \textbf{Mode 2:} $\rho(1450)\to\rho(\pi\pi)_{S\text{-wave}}$ [or $\rho\sigma$]. Here, the intermediate $\rho$ directly inherits the parent spin, which is then transmitted through the $\rho\to\pi\pi$ decay.
    \item \textbf{Mode 3:} $\rho(1450)\to\pi(1300)\pi$. Both daughters are pseudoscalars, so the parent’s spin $J=1$ must be carried entirely by the relative orbital angular momentum of the $\pi(1300)\pi$ system. The $\pi(1300)$ subsequently decays via $\pi(1300)\to\rho\pi$, but the essential spin-transfer step occurs at the first two-body decay.
\end{itemize}

These differences in intermediate quantum numbers and angular-momentum couplings lead to measurably distinct azimuthal-modulation patterns in the four-pion final state. 

Following the theoretical analysis of Refs.~\cite{Zha:2018jin,Zha:2020cst,Xing:2020hwh}, the azimuthal angle $\phi$ is defined as the angle between the transverse spin vector of the vector meson and its transverse momentum. Operationally, in this work $\phi$ is taken as the angle in the transverse plane between the reconstructed $\rho(1450)$ momentum and the momentum of the pion originating from the intermediate $\rho$ decay, evaluated in the mother’s rest frame. This choice renders the observable maximally sensitive to the spin-transfer dynamics of each decay chain.

Because the spin-transfer structure differs for the four channels, the resulting $\cos 2\phi$ modulation provides a clear discriminator among them. The measured modulation can be expressed as a weighted combination of the modulation strengths from each decay mode and reconstruction category, with weights determined by the corresponding branching fractions, fractions of correctly or incorrectly reconstructed pairs, and their intrinsic modulation responses. Consequently, the azimuthal modulation serves not only as a probe of the decay-chain composition of the $\rho(1450)$ but also as a tool for constraining the underlying quantum numbers of intermediate resonances.

\begin{table*}[!htb]
\centering
\caption{Decay modes of $\rho(1450)$}
\label{tab:rho1450_decay_modes}
\renewcommand{\arraystretch}{1.25}
\setlength{\tabcolsep}{5pt}
\begin{tabular}{c l l}
\toprule
Mode & Decay chain & $J^{PC}$ \\
\midrule

1 &
\begin{tabular}[t]{@{}l@{}}
$\rho(1450)\to a_1(1260)\pi$ \\
\phantom{$\rho(1450)\to{}$}\elbowarrow\ $a_1(1260)\to\rho\pi$ \\
\phantom{$\rho(1450)\to{}$}\phantom{\elbowarrow\ $a_1(1260)\to{}$}\elbowarrow\ $\rho\to\pi^+\pi^-$
\end{tabular}
&
\begin{tabular}[t]{@{}l@{}}
$1^{--}\to 1^{++}+0^{-}$ \\
$1^{++}\to 1^{--}+0^{-}$ \\
$1^{--}\to 0^{-}+0^{-}$
\end{tabular}
\\

2 &
\begin{tabular}[t]{@{}l@{}}
$\rho(1450)\to \rho(\pi\pi)_{S\text{-wave}}$ \\
\phantom{$\rho(1450)\to{}$}\elbowarrow\ $\rho\to\pi^+\pi^-$
\end{tabular}
&
\begin{tabular}[t]{@{}l@{}}
$1^{--}\to 1^{--}+0^{++}$ \\
$1^{--}\to 0^{-}+0^{-}$
\end{tabular}
\\

3 &
\begin{tabular}[t]{@{}l@{}}
$\rho(1450)\to \pi(1300)\pi$ \\
\phantom{$\rho(1450)\to{}$}\elbowarrow\ $\pi(1300)\to\rho\pi$ \\
\phantom{$\rho(1450)\to{}$}\phantom{\elbowarrow\ $\pi(1300)\to{}$}\elbowarrow\ $\rho\to\pi^+\pi^-$
\end{tabular}
&
\begin{tabular}[t]{@{}l@{}}
$1^{--}\to 0^{-+}+0^{-}$ \\
$0^{-+}\to 1^{--}+0^{-}$ \\
$1^{--}\to 0^{-}+0^{-}$
\end{tabular}
\\

4 &
\begin{tabular}[t]{@{}l@{}}
$\rho(1450)\to h_1(1170)\pi$ \\
\phantom{$\rho(1450)\to{}$}\elbowarrow\ $h_1(1170)\to\rho\pi$ \\
\phantom{$\rho(1450)\to{}$}\phantom{\elbowarrow\ $h_1(1170)\to{}$}\elbowarrow\ $\rho\to\pi^+\pi^-$
\end{tabular}
&
\begin{tabular}[t]{@{}l@{}}
$1^{--}\to 1^{+-}+0^{-+}$ \\
$1^{+-}\to 1^{--}+0^{-}$ \\
$1^{--}\to 0^{-}+0^{-}$
\end{tabular}
\\

5 &
$\rho(1450)\to\rho\rho$
&
$1^{--}\to 1^{--}+1^{--}$
\\

\bottomrule
\end{tabular}
\end{table*}

\section{Methods}

To quantify the spin-interference response of the different $\rho(1450)$ decay chains in the $\pi^{+}\pi^{-}\pi^{+}\pi^{-}$ final state, we perform a Monte Carlo simulation of the full cascade decays, incorporating realistic mass distributions, decay kinematics, and polarization-dependent angular weights. The event generation is structured to isolate the intrinsic differences in azimuthal modulation arising from the quantum numbers and angular-momentum couplings of the intermediate states.

\begin{figure}[H]
\centering
\includegraphics[width=0.48\textwidth]{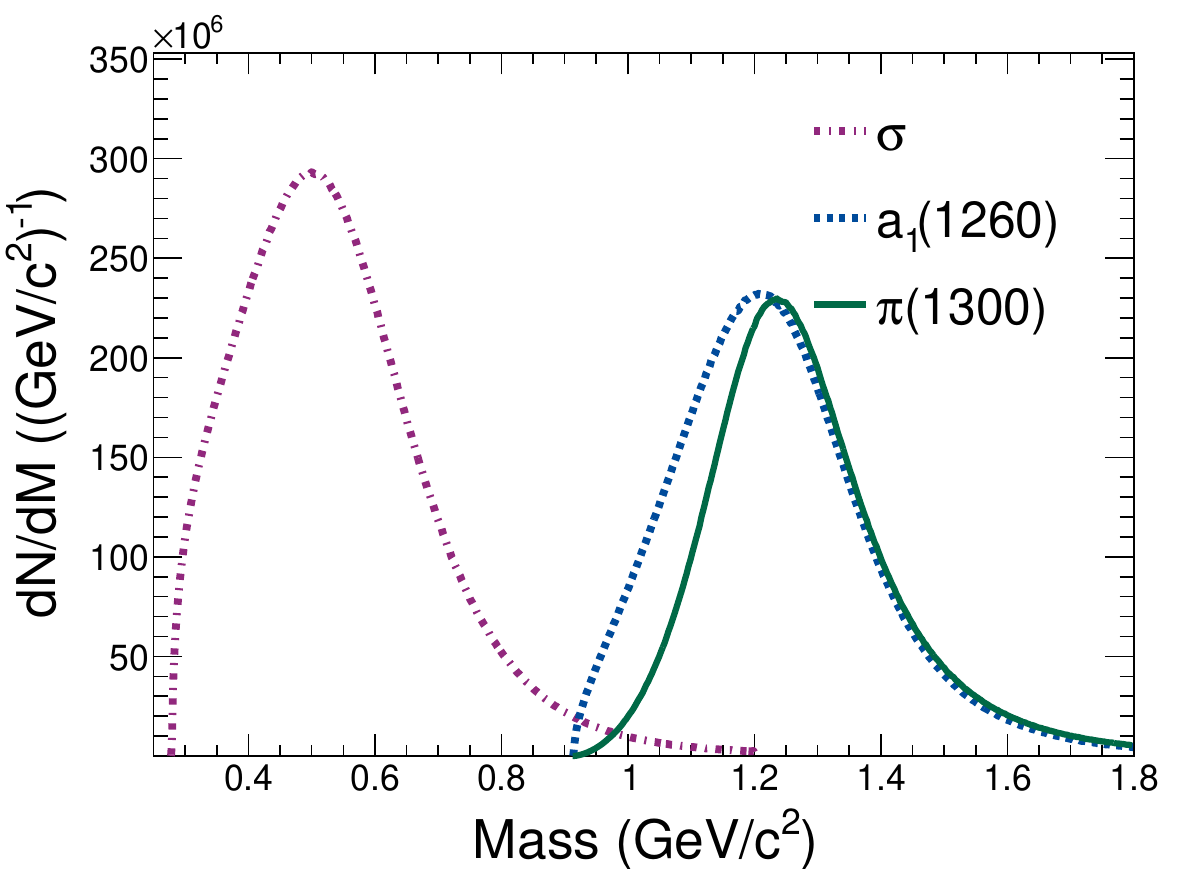}
\caption{Reconstructed invariant-mass distributions of the intermediate
resonances $a_{1}(1260)$, $\pi(1300)$, and the $(\pi\pi)_{S}$
subsystem, generated using relativistic Breit--Wigner line shapes with
mass-dependent widths.}
\label{fig:mass_distributions}
\end{figure}

\subsection{Resonance Line Shapes and Mass-Dependent Widths}

The mass distributions of the parent $\rho(1450)$ and intermediate resonances $a_{1}(1260)$, $\pi(1300)$, and the scalar $(\pi\pi)_{S}$ subsystem are sampled from relativistic Breit--Wigner line shapes with mass-dependent widths. The corresponding mass distributions of the intermediate resonances and the scalar subsystem are shown in Fig.~\ref{fig:mass_distributions}. The propagator of a resonance with pole mass $m_{0}$ and nominal width $\Gamma_{0}$ is modeled as
\begin{equation}
P(m) = \frac{1}{m_{0}^{2}-m^{2}- i m_{0}\,\Gamma(m)},
\label{eq:BW}
\end{equation}
with
\begin{equation}
\Gamma(m) = \Gamma_{0}\left(\frac{q(m)}{q(m_{0})}\right)^{2L+1}
\left(\frac{m_{0}}{m}\right)
\left[\frac{F_{L}(q(m))}{F_{L}(q(m_{0}))}\right]^{2},
\label{eq:width}
\end{equation}
where $L$ is the relative orbital angular momentum of the daughters and $q(m)$ is their breakup momentum. Blatt--Weisskopf barrier factors $F_{L}$ are implemented in the standard form~\cite{BESIII:2021xox}. This parameterization is widely used in hadronic-amplitude analyses and captures threshold and finite-size effects of the decaying system.

\subsection{Decay Dynamics for the Four Channels}

Each decay chain is generated according to the experimentally motivated dynamics of its intermediate states (see Table~\ref{tab:rho1450_decay_modes}). For the spinful intermediates in Modes~1 and 2 ($a_{1}$ and $\rho$), the polarization information is introduced at the last $\rho\to\pi\pi$ step, consistent with $s$-channel helicity conservation in photonuclear production~\cite{Gilman:1970vi,Harari:1970fw,Bialas:1971xk}. 

In Mode~3 ($\pi(1300)\pi$), both daughters of the initial decay are pseudoscalars, so the spin $J=1$ of the parent $\rho(1450)$ must be carried entirely by the orbital angular momentum of the $\pi(1300)\pi$ system. The polarization-dependent angular weight is therefore applied at the first two-body decay step, while the subsequent $\pi(1300)\to\rho\pi$ and $\rho\to\pi\pi$ decays are generated isotropically in phase space.

For all channels, the azimuthal dependence is introduced through a polarization weight of the form
\begin{equation}
w(\phi,p_{T}) \propto 1 + \alpha(p_{T})\cos 2\phi,
\label{eq:weight}
\end{equation}
where $\alpha(p_{T})$ is a common input modulation strength applied across all decay chains to ensure that differences in the reconstructed signals arise solely from the decay dynamics.

\subsection{Reconstruction of the Intermediate $\rho$ Candidate}

In each generated four-pion event, six possible pion pairs can be formed, four of which are $\pi^{+}\pi^{-}$ combinations. Following experimental practice~\cite{STAR:2009giy}, we identify the intermediate $\rho(770)$ candidate as the $\pi^{+}\pi^{-}$ pair with invariant mass closest to the $\rho$ pole mass, subject to the selection window $0.65 < m_{\pi^{+}\pi^{-}} < 0.90~\mathrm{GeV}/c^{2}$. 

Using Monte Carlo truth information, the selected pair can be categorized as:
\begin{itemize}
    \item \textit{true}: both pions originate from the intermediate $\rho$,
    \item \textit{cross}: only one pion originates from the intermediate $\rho$,
    \item \textit{recoil}: neither pion originates from the intermediate $\rho$.
\end{itemize}
These categories dilute the intrinsic modulation, and their relative fractions $f_{ij}$ depend on \pT and the decay mode.

\subsection{Construction of the Azimuthal Observable}

For each event, we compute the azimuthal angle $\phi$ between (i) the transverse momentum of the reconstructed $\rho(1450)$ and (ii) the momentum of the pion from the intermediate $\rho$ decay, evaluated in the mother’s rest frame. This definition follows Refs.~\cite{Xing:2020hwh,Zha:2020cst} and maximizes sensitivity to the spin-transfer dynamics of the decay chain.

The intrinsic modulation from decay mode $i$ and reconstruction class $j$ is quantified by
\begin{equation}
A_{ij} \equiv 2\langle \cos 2\phi\rangle_{ij}.
\label{eq:Aij}
\end{equation}
The experimentally observed modulation is a weighted sum over decay channels and reconstruction categories:
\begin{equation}
A_{\mathrm{obs}}
= \sum_{i}\sum_{j}B_{i} f_{ij}\,A_{ij},
\label{eq:Aobs_final}
\end{equation}
where $B_{i}$ denotes the branching fraction of mode~$i$ into the four-pion final state. Equation~(\ref{eq:Aobs_final}) forms the basis for extracting the $\pi(1300)\pi$ contribution from experimental measurements of $A_{\mathrm{obs}}$.

This framework allows a consistent comparison of the azimuthal-modulation responses arising from the different decay chains and enables the identification of modes, such as $\pi(1300)\pi$, that produce distinct interference signatures.

\section{Results}

\begin{figure*}[!htb]
\centering
\includegraphics[width=0.999\textwidth]{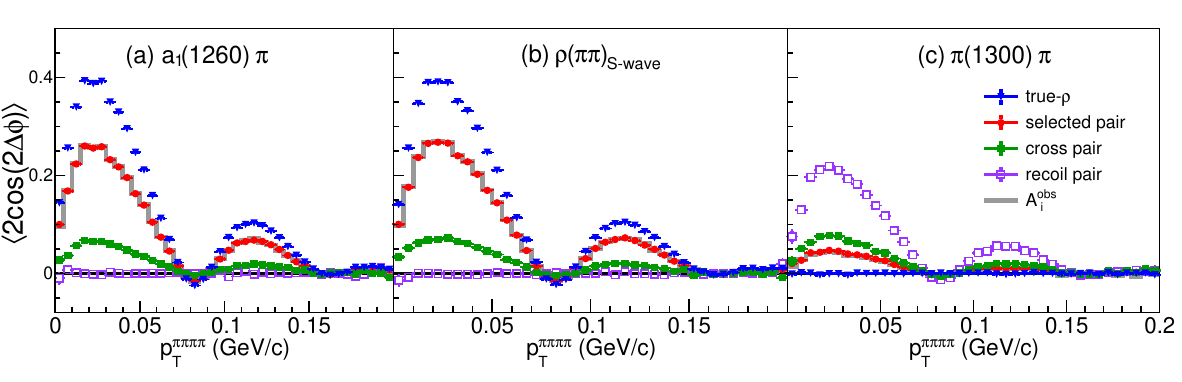}
\caption{Azimuthal-modulation observable $\langle 2\cos(2\Delta\phi)\rangle$ as a function of 
\pT for the four $\rho(1450)\rightarrow 4\pi$ decay modes: 
(a) $a_{1}(1260)\pi$,  
(b) $\rho(\pi\pi)_{S\text{-wave}}$, 
and (c) $\pi(1300)\pi$. Here, $\Delta\phi$ is the angle in the transverse plane between the reconstructed $\rho(1450)$ momentum and the momentum of the pion from the intermediate $\rho$ decay, with the pion momentum evaluated in the mother's rest frame. The red points show the modulation extracted from the experimentally motivated selected-pair reconstruction, while the blue, green, and purple points denote the true, cross, and recoil pair categories from Monte Carlo truth information. The differences among the modes reflect the distinct intermediate-state quantum numbers and angular-momentum couplings that govern how the parent-state spin alignment is transmitted to the final-state pions.}
\label{cos(2Deltaphi)_p_T}
\end{figure*}

To construct an azimuthal-modulation observable associated with the intermediate $\rho$ decay in the $\pi^{+}\pi^{-}\pi^{+}\pi^{-}$ final state, all $\pi^{+}\pi^{-}$ combinations in each event are first reconstructed. Among the four $\pi^{+}\pi^{-}$ combinations, the pair satisfying 
$0.65 < m_{\pi^{+}\pi^{-}} < 0.90~{\rm GeV}/c^{2}$ 
and having an invariant mass closest to the $\rho$ mass peak, 
$m_{\rho}\simeq 0.770~{\rm GeV}/c^{2}$, 
is selected as the experimental $\rho$ candidate. Based on this candidate, the distribution of 
$\langle 2\cos(2\Delta\phi)\rangle$ 
as a function of \pT is constructed. Figures~\ref{cos(2Deltaphi)_p_T}(a)--(c) show the \pT dependence of this observable for the three decay modes: 
$a_{1}\pi$, $\rho(\pi\pi)_{S}$, and $\pi(1300)\pi$.

The red points represent the selected-pair results obtained using the reconstruction procedure. Since the true origin of each final-state pion is known in simulation, the selected $\pi^{+}\pi^{-}$ pair can be categorized as  
(i) true (both daughters from the intermediate $\rho$),  
(ii) recoil (neither daughter from the intermediate $\rho$), or  
(iii) cross (one from the $\rho$, one from elsewhere).  
These categories reflect the combinatorial structure typical of multipion final states and have been shown to dilute spin-sensitive observables in UPC $\rho^{0}$ measurements~\cite{STAR:2022wfe}.

As seen in Fig.~\ref{cos(2Deltaphi)_p_T}, for each mode the selected-pair value of $\langle 2\cos(2\Delta\phi)\rangle$ is significantly smaller than the true-pair result. This reduction arises from cross‑pair and recoil‑pair contamination, which dilutes the azimuthal-modulation signal originally carried by the $\rho\to\pi^{+}\pi^{-}$ decay.

Because the Monte Carlo simulation tracks each pion’s origin, the fraction of reconstruction categories $f_{ij}(p_{T})$ is known for each \pT bin. Combined with the intrinsic responses $A_{ij}$, the observable for decay mode $i$ is given by
\begin{equation}
A_{i}^{\rm obs}(p_{T})
=\sum_{j} f_{ij}(p_{T})\,A_{ij}(p_{T}),
\end{equation}
in accordance with the decomposition formalism in Sec.~II.  
Figure~\ref{cos(2Deltaphi)_p_T} compares $A_{i}^{\rm obs}(p_{T})$ with the selected-pair result; the two agree to within statistical uncertainties for all modes, confirming that the observed azimuthal modulation is well described by a linear superposition of reconstruction components, as expected from the interference structure of coherent photoproduction~\cite{Xing:2020hwh,Zha:2020cst}.

\begin{figure}[!htb]
\centering
\includegraphics[width=0.48\textwidth]{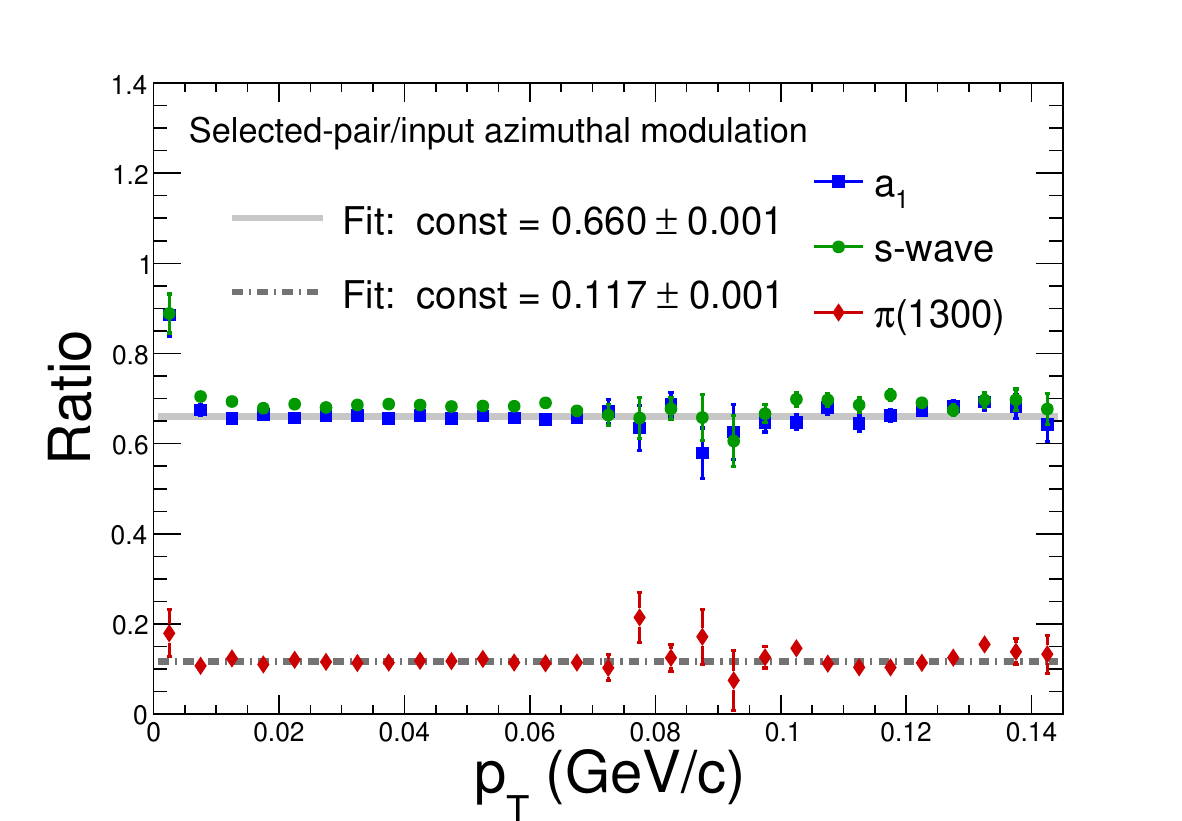}
\caption{Ratio of selected-pair to input azimuthal modulation as a
function of \pT for all decay modes, illustrating the fraction of the input modulation retained after reconstruction.}
\label{fig:Ratio}
\end{figure}

Figure~\ref{fig:Ratio} shows the ratio of selected to input modulation strength. For the $a_{1}$ and $S$‑wave channels, this ratio remains $\sim 0.660\pm0.001$ over most of the coherent region, indicating that reconstruction acts primarily as an overall normalization factor. In contrast, the $\pi(1300)\pi$ channel exhibits a substantially lower and more fluctuating ratio, reflecting its unique spin‑orbit structure and the reduced correlation between the reconstructed and true decay topology.

To compare the modulation strength of different decay modes, we focus on the region  
$0.01 < p_{T} < 0.06~{\rm GeV}/c$,  
where coherent production dominates and both statistical precision and intrinsic modulation are highest~\cite{ALICE:2022zso,ATLAS:2022yad,CMS:2020skx}. In this interval, $A_{i}^{\rm obs}$ for the  
$a_{1}\pi$ and $\rho(\pi\pi)_{S}$ modes are nearly identical. By contrast, the $\pi(1300)\pi$ mode yields a clearly distinct value, consistent with the theoretical expectation that orbital‑momentum–driven spin transfer produces a different interference pattern~\cite{Gilman:1970vi,Harari:1970fw,Bialas:1971xk}.

Motivated by this observation, the two similar modes, whose first decay step contains a spin-1 intermediate state accompanied by a spin-0 particle or system, are grouped into an effective ``Other'' component with modulation $A_{\rm Other}$. Let  
$B \equiv \mathcal{B}_{\pi(1300)\pi}$  
denote the relative branching fraction of the $\pi(1300)\pi$ channel. The observed modulation can then be written as
\begin{equation}
A_{\rm obs}
= B\,A_{\pi(1300)\pi} + (1-B)\,A_{\rm Other}.
\end{equation}
Thus, once the experimentally measured modulation is obtained, the branching fraction $B$ follows from
\begin{equation}
B=
\frac{
\langle 2\cos(2\Delta\phi)\rangle_{\rm exp}
-
\langle 2\cos(2\Delta\phi)\rangle_{\rm Other}
}{
\langle 2\cos(2\Delta\phi)\rangle_{\pi(1300)\pi}
-
\langle 2\cos(2\Delta\phi)\rangle_{\rm Other}
}.
\label{eq:a1BR}
\end{equation}

More generally, the results demonstrate that the $\phi$ distribution and its $\cos 2\phi$ modulation in $\rho(1450)\rightarrow 4\pi$ decays provide a selective probe of decay dynamics. Because each decay chain transmits the parent spin through intermediate states with distinct quantum numbers and angular-momentum couplings, the resulting azimuthal-modulation patterns differ in predictable ways. This sensitivity can be used to constrain both the relative decay fractions and, in certain cases, the quantum-number assignments of intermediate resonances in photonuclear production.

\section{Discussion}

The results above show that spin interference in coherent photonuclear production provides a sensitive and selective probe of hadronic decay dynamics. Although demonstrated using the $\rho(1450)\!\rightarrow\!4\pi$ channel, the mechanism is general: different decay chains transmit the parent vector-meson polarization through distinct combinations of intermediate-state quantum numbers and orbital-angular-momentum (OAM) couplings. These differences yield characteristic azimuthal modulations that remain visible after realistic reconstruction.

A key finding is the clear separation between the $\pi(1300)\pi$ modulation and those of the $a_{1}\pi$ and $\rho(\pi\pi)_{S}$ channels. This distinction follows from their fundamentally different spin-transfer topologies: the $\pi(1300)\pi$ mode carries polarization solely through pseudoscalar orbital motion, whereas the others involve spinful intermediates. Because this difference arises directly from angular-momentum conservation, the resulting signatures are robust against variations in line-shape modeling. The observed modulations are consistent with a linear superposition of reconstruction categories~\cite{STAR:2022wfe}. This factorization allows theoretical decay-chain responses to be mapped directly onto experimentally accessible observables. The method is broadly applicable. Many multihadron final states---including $K\bar{K}\pi$, $3\pi$, and $\pi\pi\eta$---contain mixtures of scalar, vector, and axial-vector intermediates with uncertain quantum numbers. Likewise, higher excited $\rho$ states such as $\rho(1700)$ and $\rho(2150)$~\cite{ALICE:2024kjy,STAR:2009giy} have competing interpretations that predict distinct interference patterns. Ultra-peripheral collisions (UPCs), with their well-defined photon polarization set by the impact-parameter direction, offer a uniquely clean environment to access these spin-dependent effects. Finally, Eq.~(\ref{eq:a1BR}) illustrates that interference-sensitive observables can be used to extract branching fractions, even for broad or overlapping channels where conventional amplitude analyses are difficult. The clean separation of the $\pi(1300)\pi$ response demonstrates the practical utility of this approach. 

Overall, spin-interference measurements in UPCs provide a complementary and comparatively model-independent tool for probing intermediate-state quantum numbers, with strong potential as future high-statistics UPC datasets become available.

\section{Conclusions}

We have demonstrated that production-site entanglement in coherent photonuclear interactions provides a powerful means of accessing the internal dynamics of hadronic resonances. In the $\rho(1450)\!\rightarrow\!4\pi$ system, the $a_{1}\pi$, $\rho(\pi\pi)_{S}$, and $\pi(1300)\pi$ decay chains imprint distinct entanglement-enabled $\cos 2\phi$ modulations arising from their characteristic patterns of spin transfer and orbital-angular-momentum coupling. These signatures persist after realistic reconstruction effects are included, allowing the $\pi(1300)\pi$ contribution to be cleanly isolated.

By showing that the observed modulation is a linear superposition of reconstruction components, we established a practical framework for extracting the relative branching fraction of the $\pi(1300)\pi$ mode directly from spin-interference measurements. More broadly, the method provides a robust strategy for disentangling overlapping decay channels in complex final states. It uses the coherence, controlled polarization, and long-range production-site entanglement inherent to UPCs, making it applicable to a wide class of resonances, including higher excited $\rho$ states, strange mesons, and potential exotic candidates. As higher-luminosity UPC datasets become available at RHIC, the LHC, and future facilities, entanglement-enabled spin interference can serve as a sensitive and selective tool for mapping the spectrum, internal structure, and quantum-number content of hadronic excitations.

\section{Acknowledgment}
The authors thank Professors Zuotang Liang and Wangmei Zha for the discussions. 
This work was supported in part by the National Natural Science Foundation of China under Grant No. 12475142, the Office of Nuclear Physics within the U.S. Department of Energy Office of Science under Contract DE‑FG02‑89ER40531, and the Shandong Provincial Natural Science Foundation under Grant No. ZR2024QA192.

\bibliography{reference}

\end{document}